\documentclass[twocolumn,showpacs,aps,prl,superscriptaddress]{revtex4}

\usepackage{graphicx}
\usepackage{dcolumn}
\usepackage{amsmath}
\usepackage{epsfig}

\RequirePackage{xspace}

\usepackage{relsize}
\def\babar{\mbox{\slshape B\kern-0.1em{\smaller A}\kern-0.1em
    B\kern-0.1em{\smaller A\kern-0.2em R}}}

\def\epem       {\ensuremath{e^+e^-}\xspace}

\def\pip   {\ensuremath{\pi^+}\xspace}
\def\pim   {\ensuremath{\pi^-}\xspace}

\def\Kbar  {\kern 0.2em\overline{\kern -0.2em K}{}\xspace}

\def\Kz    {\ensuremath{K^0}\xspace}
\def\Kzb   {\ensuremath{\Kbar^0}\xspace}
\def\KzKzb {\ensuremath{\Kz \kern -0.16em \Kzb}\xspace}
\def\Kp    {\ensuremath{K^+}\xspace}
\def\Km    {\ensuremath{K^-}\xspace}

\def\KpKm  {\ensuremath{\Kp \kern -0.16em \Km}\xspace}

\def\Dbar    {\kern 0.2em\overline{\kern -0.2em D}{}\xspace}

\def\Dz      {\ensuremath{D^0}\xspace}
\def\Dzb     {\ensuremath{\Dbar^0}\xspace}
\def\DzDzb   {\ensuremath{\Dz {\kern -0.16em \Dzb}}\xspace}
\def\Dp      {\ensuremath{D^+}\xspace}
\def\Dm      {\ensuremath{D^-}\xspace}

\def\DpDm    {\ensuremath{\Dp {\kern -0.16em \Dm}}\xspace}

\def\B       {\ensuremath{B}\xspace}
\def\Bbar    {\kern 0.18em\overline{\kern -0.18em B}{}\xspace}

\def\BB      {\ensuremath{B\Bbar}\xspace} 
\def\Bz      {\ensuremath{B^0}\xspace}
\def\Bzb     {\ensuremath{\Bbar^0}\xspace}
\def\BzBzb   {\ensuremath{\Bz {\kern -0.16em \Bzb}}\xspace}
\def\Bu      {\ensuremath{B^+}\xspace}
\def\Bub     {\ensuremath{B^-}\xspace}

\def\BpBm    {\ensuremath{\Bu {\kern -0.16em \Bub}}\xspace}

\mathchardef\Upsilon="7107
\def\Y#1S{\ensuremath{\Upsilon{(#1S)}}\xspace}

\def\FourS {\Y4S}

\mathchardef\Deltares="7101
\mathchardef\Xi="7104
\mathchardef\Lambda="7103
\mathchardef\Sigma="7106
\mathchardef\Omega="710A

\def\Deltabar{\kern 0.25em\overline{\kern -0.25em \Deltares}{}\xspace}
\def\Lbar{\kern 0.2em\overline{\kern -0.2em\Lambda\kern 0.05em}\kern-0.05em{}\xspace}
\def\Sigbar{\kern 0.2em\overline{\kern -0.2em \Sigma}{}\xspace}
\def\Xibar{\kern 0.2em\overline{\kern -0.2em \Xi}{}\xspace}
\def\Obar{\kern 0.2em\overline{\kern -0.2em \Omega}{}\xspace}
\def\Nbar{\kern 0.2em\overline{\kern -0.2em N}{}\xspace}
\def\Xb{\kern 0.2em\overline{\kern -0.2em X}{}\xspace}

\def\BR         {{\ensuremath{\cal B}\xspace}}

\def\upsbzbz {\ensuremath{\FourS \to \BzBzb}\xspace}

\def\mes        {\mbox{$m_{\rm ES}$}\xspace}

\newcommand{\tev}{\ensuremath{\mathrm{\,Te\kern -0.1em V}}\xspace}
\newcommand{\gev}{\ensuremath{\mathrm{\,Ge\kern -0.1em V}}\xspace}
\newcommand{\mev}{\ensuremath{\mathrm{\,Me\kern -0.1em V}}\xspace}
\newcommand{\kev}{\ensuremath{\mathrm{\,ke\kern -0.1em V}}\xspace}
\newcommand{\ev}{\ensuremath{\mathrm{\,e\kern -0.1em V}}\xspace}
\newcommand{\gevc}{\ensuremath{{\mathrm{\,Ge\kern -0.1em V\!/}c}}\xspace}
\newcommand{\mevc}{\ensuremath{{\mathrm{\,Me\kern -0.1em V\!/}c}}\xspace}
\newcommand{\gevcc}{\ensuremath{{\mathrm{\,Ge\kern -0.1em V\!/}c^2}}\xspace}
\newcommand{\mevcc}{\ensuremath{{\mathrm{\,Me\kern -0.1em V\!/}c^2}}\xspace}

\def\mus  {\ensuremath{\rm \,\mus}\xspace}

\def\ps   {\ensuremath{\rm \,ps}\xspace}

\def\mus        {\ensuremath{\,\mu{\rm s}}\xspace}    
\def\ps         {\ensuremath{{\rm \,ps}}\xspace}  

\def\to                 {\ensuremath{\rightarrow}\xspace}

\def\pep2{PEP-II}

\def\gsim{{~\raise.15em\hbox{$>$}\kern-.85em
          \lower.35em\hbox{$\sim$}~}\xspace}
\def\lsim{{~\raise.15em\hbox{$<$}\kern-.85em
          \lower.35em\hbox{$\sim$}~}\xspace}

\def\CP                {\ensuremath{C\!P}\xspace}

\def\stwoa{\ensuremath{\sin\! 2 \alpha  }\xspace}
\def\stwob{\ensuremath{\sin\! 2 \beta   }\xspace}

\def\deltat{\ensuremath{{\rm \Delta}t}\xspace}
\def\deltamd{\ensuremath{{\rm \Delta}m_d}\xspace}

\xspace

\newcommand{\jprlBase}       {Phys.\ Rev.\ Lett.\xspace}
\newcommand{\jprBase}        {Phys.\ Rev.\xspace}
\newcommand{\jplBase}        {Phys.\ Lett.\xspace}

\newcommand{\nimBaseC}       {Nucl.\ Instr.\ and Methods\xspace}

\newcommand{\npBase}         {Nucl.\ Phys.\xspace}

\newcommand{\arnps}     [1]  {{Ann.\ Rev.\ Nucl.\ Part.\ Sci.\ {\bf #1}}}

\newcommand{\nima}      [1]  {\nimBaseC~A~{\bf #1}}

\newcommand{\npb}       [1]  {\npBase\ B~{\bf #1}}

\newcommand{\plb}       [1]  {\jplBase\ B~{\bf #1}}

\newcommand{\jprl}      [1]  {\jprlBase\ {\bf #1}}
\newcommand{\jprd}      [1]  {\jprBase\ D~{\bf #1}}

\newcommand{\progtp}    [1]  {{Prog.\ Th.\ Phys.\ {\bf #1}}}

\def\jetset74   {\mbox{\tt Jetset \hspace{-0.5em}7.\hspace{-0.2em}4}\xspace}

\def\hh     {\ensuremath{h^+h^{\prime -}}}

\def\fpm {\ensuremath{f_{\pm}(\deltat)}}
\def\ilam {\ensuremath{{\cal I}m\lambda}}
\def\alam {\ensuremath{\left|\lambda\right|}}

\def\spipi {\ensuremath{S_{\pi\pi}}}
\def\cpipi {\ensuremath{C_{\pi\pi}}}
\def\de {\ensuremath{\Delta E}}

\def\Btag {\ensuremath{B_{\rm tag}}}

\def\Bflav {\ensuremath{B_{\rm flav}}}

\def\figurebox#1#2#3{%
    \def\arg{#3}%
    \ifx\arg\empty
    {\hfill\vbox{\hsize#2\hrule\hbox to #2{\vrule\hfill\vbox to #1{\hsize#2\vfill}\vrule}\hrule}\hfill}%
    \else
    {\hfill\epsfbox{#3}\hfill}%
    \fi}

\long\def\inst#1{\par\nobreak\kern 4pt\nobreak
    {\it #1}\par\vskip 10pt plus 3pt minus 3pt}

\begin{document}


\begin{flushleft}
\babar-PUB-02/09 \\
SLAC-PUB-9317\\
hep-ex/0207055\\[10mm]
\end{flushleft}

\title{
{
\Large \bf \boldmath Measurements of Branching Fractions and \CP-violating
Asymmetries in $\Bz\to\pip\pim,\, \Kp\pim,\, \Kp\Km$ Decays
}
}

%
\author{B.~Aubert}
\author{D.~Boutigny}
\author{J.-M.~Gaillard}
\author{A.~Hicheur}
\author{Y.~Karyotakis}
\author{J.~P.~Lees}
\author{P.~Robbe}
\author{V.~Tisserand}
\author{A.~Zghiche}
\affiliation{Laboratoire de Physique des Particules, F-74941 Annecy-le-Vieux, France }
\author{A.~Palano}
\author{A.~Pompili}
\affiliation{Universit\`a di Bari, Dipartimento di Fisica and INFN, I-70126 Bari, Italy }
% ...
\collaboration{The \babar\ Collaboration}
\noaffiliation


\date{\today}

\begin{abstract}
We present measurements of branching fractions and $\CP$-violating
asymmetries for neutral $B$ meson decays to two-body final states of
charged pions and kaons based on a sample of about $88$ million
$\Y4S\to\BB$ decays. From a time-independent fit we measure the
charge-averaged branching fractions $\BR(\Bz\to\pip\pim) = (4.7\pm
0.6\pm 0.2)\times 10^{-6}$, $\BR(\Bz\to\Kp\pim) = (17.9\pm 0.9\pm
0.7)\times 10^{-6}$, and the direct \CP-violating charge asymmetry
${\cal A}_{K\pi} = -0.102 \pm 0.050\pm
0.016\;\left[-0.188,-0.016\right]$, where the ranges in square
brackets indicate the $90\%$ confidence intervals. From a
time-dependent fit we measure the $\Bz\to\pip\pim$ \CP-violating
parameters $\spipi = 0.02\pm 0.34\pm 0.05\; \left[-0.54,+0.58\right]$
and $\cpipi = -0.30\pm 0.25\pm 0.04\; \left[-0.72,+0.12\right]$.

\end{abstract}


\maketitle

\maketitle

Recent measurements of the $\CP$-violating asymmetry parameter $\stwob$ reported
by the \babar~\cite{BaBarSin2betaObs} and Belle~\cite{BelleSin2betaObs} 
Collaborations established \CP violation in neutral $B$ decays.  
These results are consistent with the Standard Model (SM)
expectation based on indirect constraints on the 
magnitudes of the elements of the Cabibbo-Kobayashi-Maskawa~\cite{CKM} 
quark-mixing matrix.  However, a full test of the \CP\ violation mechanism
in the SM, through a single complex phase in the CKM matrix, will require
additional direct constraints on the angles ($\alpha$, $\beta$, and
$\gamma$) of the Unitarity Triangle~\cite{nirandquinn}.

The time-dependent \CP-violating asymmetry in the decay $\Bz\to\pip\pim$ 
is related to the angle $\alpha$, and ratios of 
branching fractions for various $\pi\pi$ 
and $K\pi$ decay modes are sensitive to the angle $\gamma$.  
In this Letter we present results for branching fractions and \CP-violating 
asymmetries in $\Bz\to\pip\pim$, $\Kp\pim$, and $\Kp\Km$ decays~\cite{ref:cc}
using a sample of $87.9\pm 1.0$ million $\B\Bbar$ pairs.  A detailed 
description of the \babar\ detector is presented in Ref.~\cite{ref:babar}, and
more details on the analysis technique are given in 
Refs.~\cite{OldPubs}, which describe our previous measurements
of these quantities.  Other measurements of the branching
fractions and the charge asymmetry in $\Bz\to\Kp\pim$ have been
performed by the CLEO and Belle Collaborations ~\cite{bellecleoBR}.
More recently, the Belle Collaboration reported a measurement of
the time-dependent \CP asymmetry in
\Bz\to\pip\pim~\cite{ref:BelleSin2alpha}.

We reconstruct a sample of neutral $B$ mesons ($B_{\rm rec}$) decaying to the $\hh$ 
final state, where $h$ and $h^{\prime}$ refer to $\pi$ or $K$.
Signal yields are determined with a maximum likelihood fit including kinematic,
topological, and particle identification information.
For the $K^{\mp}\pi^{\pm}$ components, the yield is parameterized as 
$N_{K^{\mp}\pi^{\pm}} = N_{K\pi}\left(1 \pm {\cal A}_{K\pi}\right)/2$, where 
$N_{K\pi}$ is the total yield and
${\cal A}_{K\pi}\equiv (N_{\Km\pip} - N_{\Kp\pim})/(N_{\Km\pip} + N_{\Kp\pim})$
is the \CP-violating charge asymmetry.
The asymmetry arises from interference between the $b\to s$ penguin
and $b\to u$ tree amplitudes, and is
predicted~\cite{ref:bbns,ref:directCP} to be less than $20\%$ in the
Standard Model.  However, a larger asymmetry could be induced by new
particles, such as charged Higgs bosons or supersymmetric particles,
contributing to the penguin amplitude.

In order to extract the \CP\ asymmetry parameters in $\Bz\to\pip\pim$
decay, we examine each event in the $B_{\rm rec}$ sample to determine
whether the second $B$ meson (\Btag) decayed as a $\Bz$ or $\Bzb$
(flavor tag) and reconstruct the difference $\deltat$ between the
proper decay times of the $B_{\rm rec}$ and \Btag\ decays.
The decay rate distribution $f_+\,(f_-)$ when $\hh = \pip\pim$ and 
$\Btag = \Bz\,(\Bzb)$ is given by
\begin{eqnarray}
\fpm = \frac{e^{-\left|\deltat\right|/\tau}}{4\tau} [1
& \pm & \spipi\sin(\deltamd\deltat) \nonumber \\
& \mp & \cpipi\cos(\deltamd\deltat)],
\label{fplusminus}
\end{eqnarray}
where $\tau$ is the mean $\Bz$ lifetime and $\deltamd$ is the mixing
frequency due to the eigenstate mass difference.
The parameters $\spipi$ and $\cpipi$ are defined as
\begin{equation}
\spipi \equiv \frac{2\,\ilam}{1+\alam^2}\quad{\rm and}\quad \cpipi \equiv \frac{1-\alam^2}{1+\alam^2},
\label{SandCdef}
\end{equation}
and vanish in the absence of $\CP$ violation.  If the decay proceeds 
purely through the $b\to u$ tree amplitude, the complex parameter
$\lambda$ is given by
\begin{equation}
\lambda(B\to\pip\pim) 
= \left(\frac{V_{\rm tb}^{*}V_{\rm td}^{}}{V_{\rm tb}^{}V_{\rm td}^{*}}\right)
\left(\frac{V_{\rm ud}^{*}V_{\rm ub}^{}}{V_{\rm ud}^{}V_{\rm ub}^{*}}\right).
\end{equation}
In this case $\cpipi = 0$ and $\spipi = \stwoa$, where 
$\alpha \equiv \arg\left[-V_{\rm td}^{}V_{\rm tb}^{*}/V_{\rm ud}^{}V_{\rm ub}^{*}\right]$.  
In general, the $b\to d$ penguin amplitude modifies both the magnitude and phase
of $\lambda$, so that $\cpipi \ne 0$ and 
$\spipi = \sqrt{1 - \cpipi^2}\sin{2\alpha_{\rm eff}}$, 
where $\alpha_{\rm eff}$ depends on the magnitudes and relative strong and weak
phases of the tree and penguin amplitudes.  Several approaches have been proposed 
to obtain information on $\alpha$ in the presence of 
penguins~\cite{ref:bbns,alphafrompenguins}.

The event selection and $B_{\rm rec}$ reconstruction used in this analysis are
similar to those used in Ref.~\cite{OldPubs}.  Hadronic events are selected
based on charged particle multiplicity and event topology.
Candidate $\B_{\rm rec}$ decays are reconstructed from pairs of oppositely-charged 
tracks forming a good quality vertex, where the $B_{\rm rec}$ four-momentum is
calculated with the pion mass assumed for both tracks.  

Signal decays are identified kinematically using two variables,
the difference $\de$ between the center-of-mass (CM) energy of the $B_{\rm rec}$ 
candidate and $\sqrt{s}/2$, and the beam-energy substituted mass
$\mes = \sqrt{(s/2 + {\mathbf {p}}_i\cdot {\mathbf {p}}_B)^2/E_i^2- {\mathbf {p}}_B^2}$, 
where $\sqrt{s}$ is the total CM energy, and the $B_{\rm rec}$ momentum ${\mathbf {p_B}}$ 
and the four-momentum of the initial state $(E_i, {\mathbf {p_i}})$ are 
defined in the laboratory frame.
For signal decays $\de$ and $\mes$ are Gaussian distributed with resolutions
of $26\mev$ and $2.6\mevcc$, respectively.  For $\pip\pim$ decays $\de$ peaks
near zero, while for decays with one or two kaons the $\de$ peak position is
parameterized as a function of the kaon momenta in the laboratory frame, with
an average shift of $-45\mev$ and $-91\mev$, respectively.
The distribution of $\mes$ peaks near the $B$ mass.
We require $5.20 < \mes < 5.29\gevcc$ and $\left|\de\right|<0.15\gev$.

Identification of $\hh$ tracks as pions or kaons is accomplished with
the Cherenkov angle measurement $\theta_c$ from a detector of
internally reflected Cherenkov light. 
 We construct charge-dependent
double-Gaussian probability density functions (PDFs) from the
difference between measured and expected values of $\theta_c$ for the
pion or kaon hypothesis, normalized by the error $\sigma_{\theta_c}$.
The PDF parameters are measured in a sample of $D^{*+}\to\Dz\pip$,
$\Dz\to \Km\pip$ decays, reconstructed in data.  The typical
separation between pions and kaons varies from $8\sigma_{\theta_c}$ at
$2\gevc$ to $2.5\sigma_{\theta_c}$ at $4\gevc$.

We have studied potential backgrounds from other $B$ decays and find
them to be negligible.  Backgrounds from the process $\epem\to
q\bar{q}\; (q=u,d,s,c)$ are suppressed by their topology.  In the CM
frame we define the angle $\theta_S$ between the sphericity axis of
the $B$ candidate and the sphericity axis of the remaining particles
in the event, and require $\left|\cos{\theta_S}\right|<0.8$, which
removes $83\%$ of this background.  
For these particles we also define a Fisher discriminant ${\cal F} =
0.53 - 0.60\times \sum_i{p_i^*} + 1.27\times
\sum_i{p^*_i\left|\cos(\theta_i^*)\right|^2}$ where $p^*_i$ is the
momentum of particle $i$ and $\theta^*_i$ is the angle between its
momentum and the $\B_{\rm rec}$ thrust axis in the CM frame.  The
shapes of ${\cal F}$ for signal and background events are included as
PDFs in the maximum likelihood fit.

We use a multivariate technique~\cite{ref:sin2betaPRL02} to determine the flavor of 
the $\Btag$ meson.  Separate neural networks are trained to identify primary leptons, kaons, 
soft pions from $D^*$ decays, and high-momentum charged particles from \B\ decays.  
Events are assigned to one of five mutually exclusive tagging categories 
based on the estimated mistag probability and the source of the tagging information 
(Table~\ref{tab:tagging}).
The quality of tagging is expressed in terms of the effective efficiency 
$Q = \sum_k \epsilon_k (1-2w_k)^2$, where $\epsilon_k$ and $w_k$ are the 
efficiencies and mistag probabilities, respectively, for events tagged in category $k$.
Table~\ref{tab:tagging} summarizes the tagging performance measured in a data sample
\Bflav\ of fully reconstructed neutral $B$ decays to 
$D^{(*)-}(\pip,\, \rho^+,\, a_1^+)$.  The assumption of equal tagging efficiencies and mistag probabilities
for signal $\pip\pim$, $\Kp\pim$, and $\Kp\Km$ decays is validated in a detailed 
Monte Carlo simulation. 
The background hypothesis have separate tagging efficiencies.

\begin{table}[!tbp]
\caption{Average tagging efficiency $\epsilon$, average mistag fraction $w$,
mistag fraction difference $\Delta w = w(\Bz) - w(\Bzb)$, and effective tagging efficiency 
$Q$ for signal events in each tagging category.  The quantities are measured in the 
\Bflav\ sample.}
\smallskip
\begin{center}
\begin{tabular}{crclrclrclrcl} \hline\hline
Category & \multicolumn{3}{c}{$\epsilon\,(\%)$} & \multicolumn{3}{c}{$w\,(\%)$} & \multicolumn{3}{c}{$\Delta w\,(\%)$} &
\multicolumn{3}{c}{$Q\,(\%)$} \rule[-2mm]{0mm}{6mm} \\\hline
{\tt Lepton}    & $9.1  $&$ \pm $&$ 0.2 $&$ 3.3  $&$ \pm $&$ 0.7 $&$ -1.6 $&$ \pm $&$ 1.3 $&$ 8.0  $&$ \pm $&$ 0.3$\\
{\tt Kaon\,I}   & $16.6 $&$ \pm $&$ 0.2 $&$ 9.5  $&$ \pm $&$ 0.7 $&$ -2.8 $&$ \pm $&$ 1.3 $&$ 10.7 $&$ \pm $&$ 0.4$\\
{\tt Kaon\,II}  & $19.8 $&$ \pm $&$ 0.3 $&$ 20.6 $&$ \pm $&$ 0.8 $&$ -5.3 $&$ \pm $&$ 1.3 $&$ 6.7  $&$ \pm $&$ 0.4$\\
{\tt Inclusive} & $20.1 $&$ \pm $&$ 0.3 $&$ 31.7 $&$ \pm $&$ 0.9 $&$ -2.6 $&$ \pm $&$ 1.4 $&$ 2.7  $&$ \pm $&$ 0.3$\\
{\tt Untagged}  & $34.4 $&$ \pm $&$ 0.5 $\\\hline
Total $Q$       &        &       &       &        &       &       &        &       &       &$ 28.4 $&$ \pm $&$ 0.7$ \rule[-2mm]{0mm}{6mm} \\\hline\hline
\end{tabular}
\end{center}
\label{tab:tagging}
\end{table}

The time difference $\deltat$ is obtained from the known boost of the
$\epem$ system and the measured distance between the $z$ positions of
the $B_{\rm rec}$ and $\Btag$ decay vertices.  A detailed description
of the algorithm is given in Ref.~\cite{ref:Sin2betaPRD}.  We require
$\left|\deltat\right|<20\ps$ and $\sigma_{\deltat} < 2.5\ps$, where
$\sigma_{\deltat}$ is the error on $\deltat$.  The resolution function
for signal candidates is a sum of three Gaussians, identical to the
one described in Ref.~\cite{ref:sin2betaPRL02}, with parameters
determined from a fit to the \Bflav\ sample (including events in all
five tagging categories).  The background $\deltat$ distribution is
modeled as the sum of an exponential convolved with a Gaussian, with
two additional Gaussians to account for tails. 
Common parameters are used to describe the background shape for all
  tagging categories.
We find that $96\%$ of background events are
described by an effective lifetime of approximately $0.7\ps$.

We use an unbinned extended maximum likelihood fit to extract yields and $\CP$ parameters
from the $B_{\rm rec}$ sample.  The likelihood for candidate $j$ tagged in category 
$k$ is obtained by summing the product of event yield $N_{i}$, tagging efficiency $\epsilon_{i,k}$,
and probability ${\cal P}_{i,k}$ over the eight possible signal and background hypotheses $i$
(referring to $\pi^{+}\pi^{-}$, $K^{+}\pi^{-}$, $K^{-}\pi^{+}$, and $K^{+}K^{-}$ decays).
The extended likelihood function for category $k$ is
\begin{equation}
{\cal L}_k = \exp{\left(-\sum_{i}N_i\epsilon_{i,k}\right)}
\prod_{j}\left[\sum_{i}N_i\epsilon_{i,k}{\cal P}_{i,k}(\vec{x}_j;\vec{\alpha}_i)\right].
\end{equation}
The probabilities ${\cal P}_{i,k}$ are evaluated as the product of PDFs 
for each of the independent variables 
$\vec{x}_j = \left\{\mes, \de, {\cal F}, \theta_c^+, \theta_c^-, \deltat\right\}$, 
where $\theta_c^+$ and $\theta_c^-$ are the Cherenkov angles for the positively and 
negatively charged tracks.  
We use separate PDF parameters 
for $\theta_c^+$ and $\theta_c^-$ to account for possible systematic differences.  
The total likelihood ${\cal L}$ is the product of likelihoods for each tagging category,
and the free parameters are determined by maximizing the quantity $\ln{\cal L}$.
The fitted sample contains $26070$ events.

Signal yields are determined from a fit excluding tagging or 
$\deltat$ information.  There are $16$ free parameters, including signal and 
background yields (6 parameters); $K\pi$ asymmetries (2); and
parameters for the background shapes in $\mes$ (1), 
$\de$ (2), and ${\cal F}$ (5).  
Table~\ref{tab:BR} summarizes signal yields, total efficiencies, 
charge-averaged branching fractions, and ${\cal A}_{K\pi}$.  
In the efficiency calculation we neglect possible effects due to final state 
radiation from the $B_{\rm rec}$ decay products.
The significance of ${\cal A}_{K\pi}$ is $2.0$, where significance is
defined as the square root of the change in $-2\log{\cal L}$ when
${\cal A}_{K\pi}$ is fixed to zero.
These results
are consistent with our previous measurements~\cite{OldPubs}, and with
measurements from other experiments~\cite{bellecleoBR}.
For the decay $\Bz\to\Kp\Km$ we measure a yield of only $1\pm 8$ events and so
compute a Bayesian $90\%$ confidence level (C.L.) upper limit on the
branching fraction.  Ref.~\cite{OldPubs} gives a detailed description of the
method used.


The dominant sources of systematic error on the branching fraction measurements are
from possible fit bias (determined in large samples of Monte Carlo simulated events),
uncertainty in track and $\theta_c$ reconstruction efficiencies,
and imperfect knowledge of the PDF shapes.  
The calculation of selection efficiencies using Monte Carlo simulated decays
has been checked against control samples in data and residual uncertainties are
included in the systematic error on branching fractions.
For ${\cal A}_{K\pi}$ the systematic error is dominated by the $\theta_c$ PDF shape
and possible charge bias in track reconstruction.  The total systematic error is 
computed as the sum in quadrature of the individual uncertainties.

\begin{table*}[!htb]
\begin{center}
\caption{Summary of results for total detection efficiencies, fitted signal 
yields $N_S$, charge-averaged branching fractions \BR, and ${\cal A}_{K\pi}$.
Branching fractions are calculated assuming equal rates for \upsbzbz\ and $\Bu\Bub$.  
The upper limits for $N_{\Kp\Km}$ and $\BR(\Bz\to\Kp\Km)$ correspond to 
the $90\%$ C.L.}
\label{tab:BR}
\begin{ruledtabular}
\begin{tabular}{lccccc} 
Mode  & Efficiency (\%) & $N_S$ & \BR($10^{-6}$) & ${\cal A}_{K\pi}$ & ${\cal A}_{K\pi}$ $90\%$ C.L. \\ 
\hline
$\pip\pim$ & $38.0\pm 0.8$  & $157\pm 19\pm 7$ & $4.7\pm 0.6\pm 0.2$ & & \\
$\Kp\pim$ & $37.5\pm 0.8$ & $589\pm 30\pm 17$ & $17.9\pm 0.9\pm 0.7$ & $-0.102\pm 0.050\pm 0.016$ & $[-0.188,-0.016]$ \\
$\Kp \Km$  & $36.2\pm 0.8$ & $1\pm 8\,(<16)$ & $<0.6$ & & \\
\end{tabular}
\end{ruledtabular}
\end{center}
\end{table*}

\begin{figure}[!tbp]
\begin{center}
\includegraphics[width=0.45\linewidth]{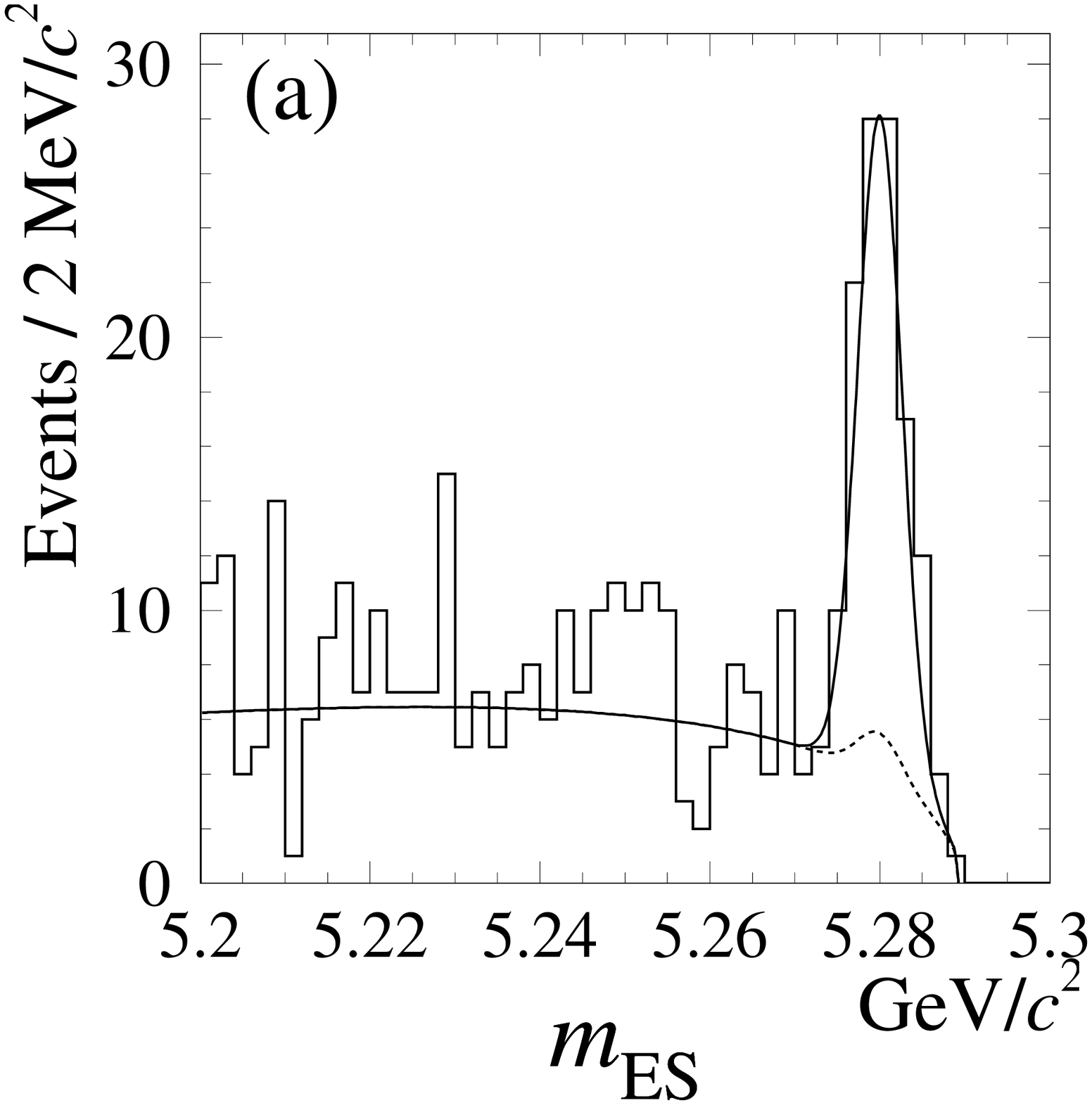}\includegraphics[width=0.45\linewidth]{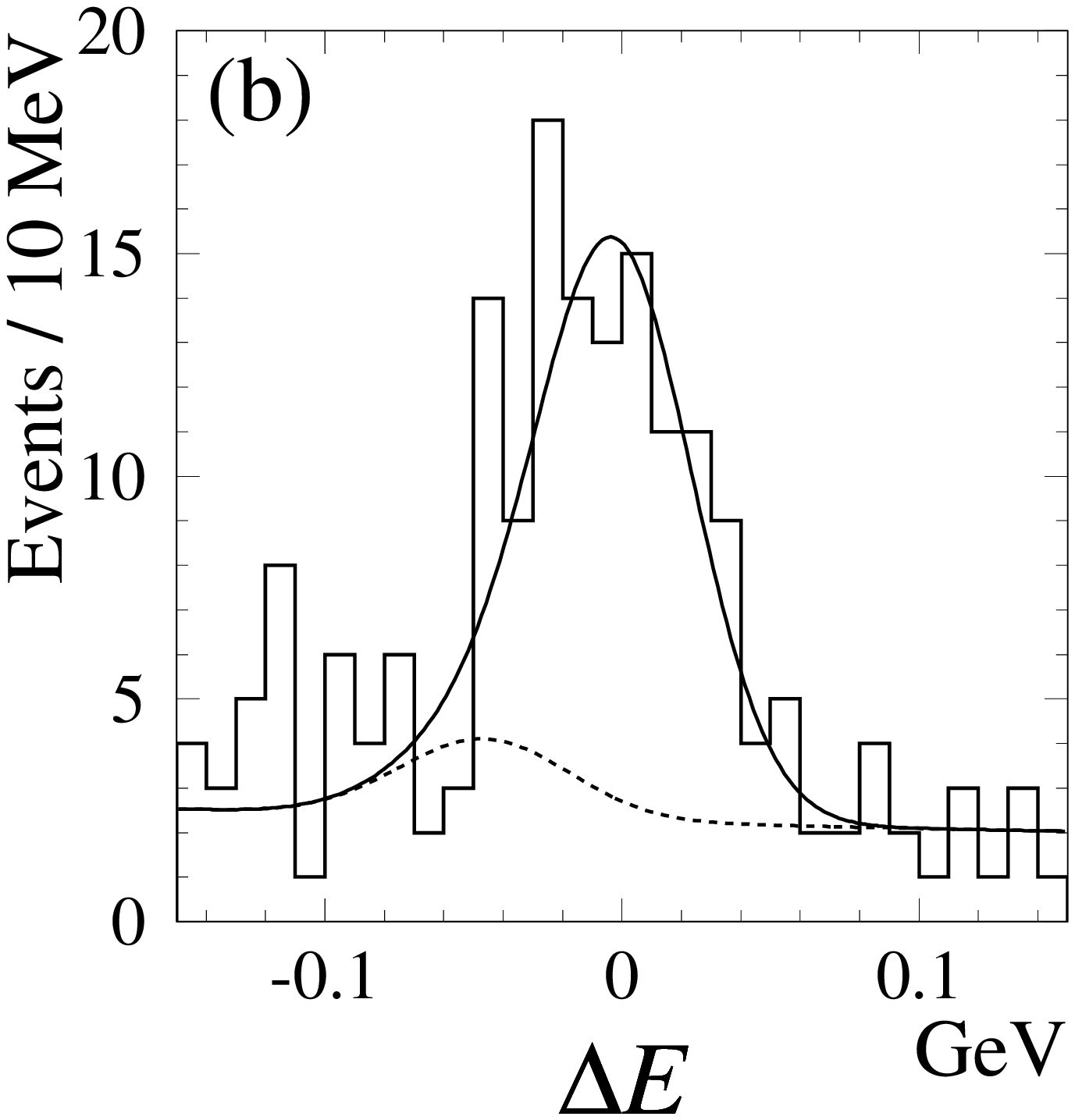}
\includegraphics[width=0.45\linewidth]{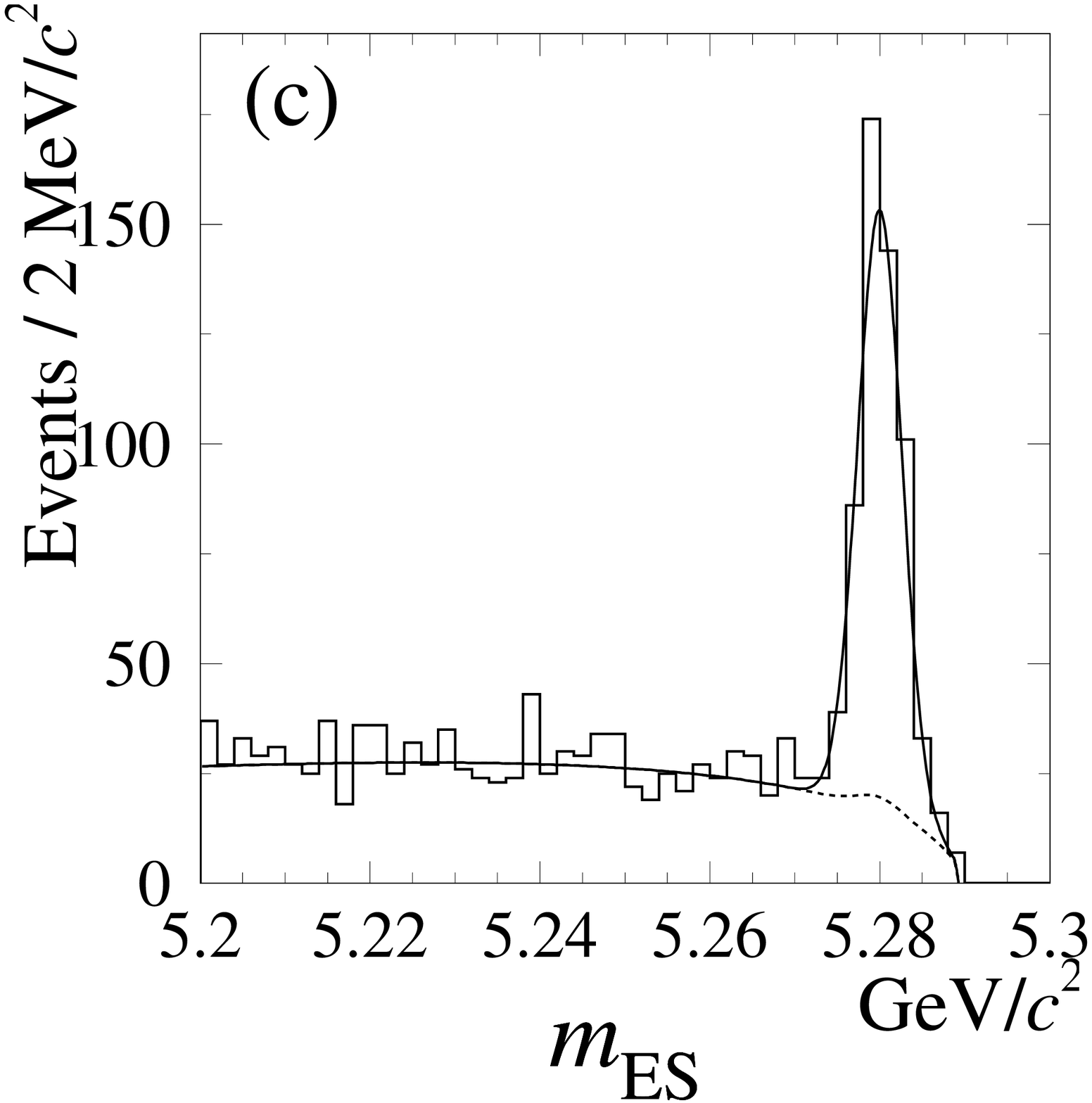}\includegraphics[width=0.45\linewidth]{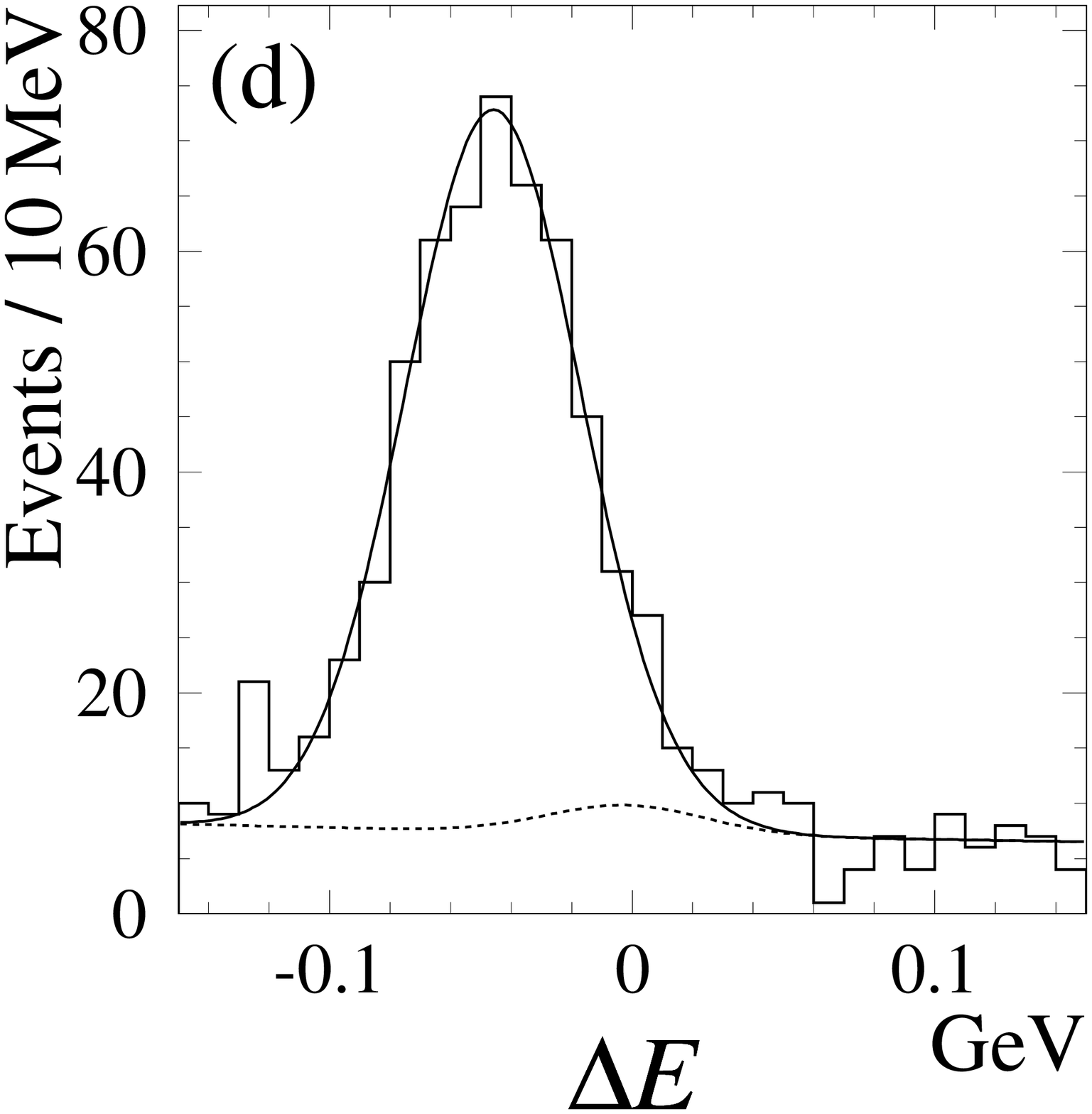}
\caption{Distributions of $\mes$ and $\de$
for events enhanced in signal (a), (b) $\pip\pim$ and (c), (d) $K^{\mp}\pi^{\pm}$
decays.  Solid curves represent projections of the maximum likelihood fit, 
dashed curves represent $q\bar{q}$ and $\pi\pi\leftrightarrow K\pi$ 
cross-feed background.}
\label{fig:prplots}
\end{center}
\end{figure}

Figure~\ref{fig:prplots} shows distributions of $\mes$ and $\de$ after selecting on 
probability ratios to enhance the signal purity.  
The solid curve in each plot represents the fit projection after correcting for the
efficiency of the additional selection ($52\%$ for $\pi\pi$, $79\%$ for $K\pi$).

The parameters $\spipi$ and $\cpipi$ are determined from a second fit 
including tagging and $\deltat$ information, where the $\Bflav$ sample is included
to determine the signal parameters describing tagging information and 
the $\deltat$ resolution function.
The $\deltat$ PDF for signal $\pip\pim$ 
decays is given by Eq.~\ref{fplusminus}, modified to include $w_k$ and $\Delta w_k$ for
each tagging category and convolved with the signal resolution function.  We also
take into account possible differences in reconstruction and tagging efficiencies
between $\Bz$ and $\Bzb$ mesons.  The $\deltat$ PDF for signal $\Kp\pim$ events 
takes into account $\Bz$--$\Bzb$ mixing based on the charge of the kaon and the flavor of 
$\Btag$.

A total of $76$ parameters are varied in the fit, including the values of $\spipi$ and
$\cpipi$ $(2)$; signal and background yields $(5)$; $K\pi$ charge asymmetries $(2)$; 
signal and background tagging efficiencies $(16)$ and 
efficiency asymmetries $(16)$; signal mistag fraction and mistag fraction 
differences $(8)$; signal resolution function $(9)$; and parameters for the 
background shapes in 
$\mes$ $(5)$, $\de$ $(2)$, ${\cal F}$ $(5)$, and $\deltat$ $(6)$.  We assume
zero events from $\Bz\to\Kp\Km$ decays and we fix 
$\tau_{\Bz}$ and $\deltamd$ to their world average values~\cite{PDG2002}.
As a means of validating the analysis technique, we determine $\tau$ and $\deltamd$ in the 
$B_{\rm rec}$ sample and find $\tau = (1.56\pm 0.07)\ps$ and 
$\deltamd = (0.52\pm 0.05)\ps^{-1}$.  

\begin{figure}[!tbp]
\begin{center}
\includegraphics[width=0.5\linewidth]{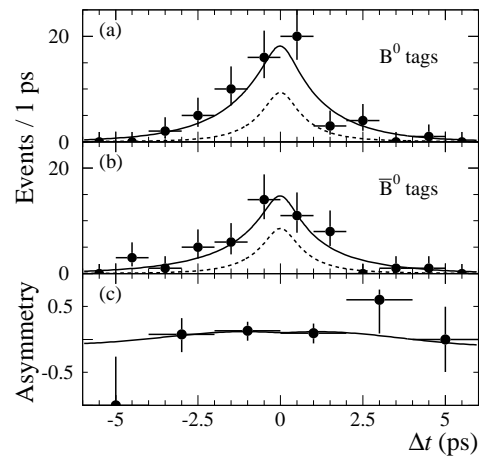}
\end{center}
\caption{
Distributions of $\deltat$ for events enhanced in 
signal $\pi\pi$ decays with $B_{\rm tag}$ tagged as
(a) $\Bz$ ($N_{\Bz}$) or (b) $\Bzb$ ($N_{\Bzb}$), and (c) the asymmetry
$\left[N_{\Bz} - N_{\Bzb}\right]/\left[N_{\Bz} + N_{\Bzb}\right]$ 
as a function of $\deltat$.  Solid curves represent 
projections of the maximum likelihood fit, dashed curves 
represent the sum of $q\bar{q}$ and $K\pi$ background events.}
\label{fig:dtplot}
\end{figure}

The combined fit to the $B_{\rm rec}$ and $B_{\rm flav}$ samples yields
\begin{eqnarray*}
\spipi & =          & 0.02\pm 0.34\,({\rm stat})\pm 0.05\,({\rm syst})\;  \left[-0.54,+0.58\right],\\
\cpipi & =          & -0.30\pm 0.25\,({\rm stat})\pm 0.04\,({\rm syst})\;  \left[-0.72,+0.12\right],
\end{eqnarray*}
where the range in square brackets indicates the $90\%$ C.L.~interval
taking into account the systematic errors.
The correlation between $\spipi$ and $\cpipi$ is $-10\%$.
The signal yields determined in this fit are within 
$3\%$ of the yields obtained from the time-independent fit.  
Systematic uncertainties on $\spipi$ and $\cpipi$ are 
dominated by imperfect knowledge of the PDF shapes and possible fit bias.  
We also evaluate multiplicative systematic errors $(0.015)$, which
are calculated at one standard deviation and summed in quadrature with the 
additive systematic uncertainties.
Figure~\ref{fig:dtplot} shows distributions of $\deltat$ for events with $B_{\rm tag}$ tagged
as $\Bz$ or $\Bzb$, and the asymmetry 
as a function of $\deltat$ for tagged events enhanced in signal $\pi\pi$ decays.

In summary, we have presented updated measurements of branching fractions and \CP-violating
asymmetries in $\Bz\to\pip\pim$, $\Kp\pim$, and $\Kp\Km$ decays.  These results are consistent 
with, and supersede our previous measurements~\cite{OldPubs}.  We do not observe
large mixing-induced or direct $\CP$ violation in the time-dependent asymmetry of 
$\Bz\to\pip\pim$ decays, as reported in~\cite{ref:BelleSin2alpha}.

\par

We are grateful for the excellent luminosity and machine conditions
provided by our \pep2\ colleagues, 
and for the substantial dedicated effort from
the computing organizations that support \babar.
The collaborating institutions wish to thank 
SLAC for its support and kind hospitality. 
This work is supported by
DOE
and NSF (USA),
NSERC (Canada),
IHEP (China),
CEA and
CNRS-IN2P3
(France),
BMBF and DFG
(Germany),
INFN (Italy),
NFR (Norway),
MIST (Russia), and
PPARC (United Kingdom). 
Individuals have received support from the 
A.~P.~Sloan Foundation, 
Research Corporation,
and Alexander von Humboldt Foundation.


\begin{thebibliography}{99}

\bibitem{BaBarSin2betaObs}
\babar\ Collaboration, B. Aubert {\em et al.}, \jprl{87}, 091801 (2001).

\bibitem{BelleSin2betaObs}
BELLE Collaboration, K. Abe {\em et al.}, \jprl{87}, 091802 (2001).

\bibitem{CKM}
N.~Cabibbo, \jprl {\bf 10}, 531 (1963);
M.~Kobayashi and T.~Maskawa, \progtp {\bf 49}, 652 (1973).

\bibitem{nirandquinn}
For a general review see Y. Nir and H. Quinn,
\arnps{42}, 211 (1992).

\bibitem{ref:cc}
Unless explicitly stated, charge conjugate decay modes are assumed throughout this paper.

\bibitem{ref:babar}
\babar\ Collaboration, B. Aubert {\em et al.}, \nima{479}, 1 (2002).

\bibitem{OldPubs}
\babar\ Collaboration, B. Aubert {\em et al.}, \jprl{87}, 151802 (2001);
\jprd{65}, 051502 (2002).

\bibitem{bellecleoBR}
CLEO Collaboration, S. Chen {\em et al.}, \jprl{85}, 525 (2000);
CLEO Collaboration, D. Cronin-Hennessy {\em et al.}, \jprl{85}, 515 (2002);
Belle Collaboration, B.C.K. Casey {\em et al.}, hep-ex/0207090, submitted
to \jprd.

\bibitem{ref:BelleSin2alpha}
Belle Collaboration, K. Abe {\em et al.}, \jprl{89}, 071801 (2002).

\bibitem{ref:bbns}
M. Beneke, G. Buchalla, M. Neubert, and C.T. Sachrajda, \npb{606}, 245 (2001);

\bibitem{ref:directCP}
A. Ali, G. Kramer, and C. L\"{u}, \jprd{59}, 0140054 (1999);
Y.Y. Keum, H-n. Li, and A.I. Sanda, \jprd{63}, 054008 (2001); 
M. Ciuchini {\em et al.}, \plb{515}, 33 (2001).

\bibitem{alphafrompenguins}
M. Gronau and D. London, \jprl{65}, 3381 (1990); 
Y. Grossman and H.R. Quinn, \jprd{58}, 017504 (1998); J. Charles, \jprd{59}, 054007 (1999);
M. Gronau, D. London, N. Sinha, and R. Sinha, \plb{514}, 315 (2001);
M. Gronau and J. Rosner, \jprd{66}, 053003 (2002);
R. Fleischer and J. Matias, \jprd{66}, 054009 (2002).

\bibitem{ref:sin2betaPRL02}
\babar\ Collaboration, B. Aubert {\em et al.}, SLAC-PUB-9293, hep-ex/0207042, 
to appear in \jprl

\bibitem{ref:Sin2betaPRD}
\babar\ Collaboration, B. Aubert {\em et al.}, \jprd{66}, 032003 (2002).

\bibitem{PDG2002}
Particle Data Group, K. Hagiwara {\em et al.}, \jprd{66}, 010001 (2002).

\end{thebibliography}
\end{document}